\documentclass[%
 reprint,
 amsmath,amssymb,
 aps,
]{revtex4}

\usepackage{graphicx}
\usepackage{dcolumn}
\usepackage{bm}

\begin{document}

\preprint{APS/123-QED}

\title{Sparse Identification for Nonlinear Optical Communication  Systems: SINO Method}

\author{Mariia Sorokina$^{*}$, Stylianos Sygletos,  and Sergei Turitsyn }

\address{Aston Institute of Photonic Technologies, Aston University, B4 7ET Birmingham UK}

\email{m.sorokina@aston.ac.uk } 

\begin{abstract}
We introduce low complexity machine learning based approach for mitigating nonlinear impairments in optical fiber communications systems. The immense intricacy of the problem calls for the development of "smart" methodology, simplifying the analysis without losing the key features that are important for recovery of transmitted data. The proposed sparse identification method for optical systems (SINO) allows to determine the minimal (optimal) number of degrees of freedom required for adaptive mitigation of detrimental nonlinear effects. We demonstrate successful application of the SINO method both for standard fiber communication links and for few-mode spatial-division-multiplexing systems.  
\end{abstract}
\maketitle


\section{Introduction}

There is an enormous pressure on the fiber-optic communication industry to deal with the exponentially increasing capacity demands from data traffic \cite{DJR,CLEO} driven by the existing and constantly emerging internet and broadband services as well as the fast growing machine-to-machine traffic. Current technological solutions suggest a combination of advanced modulation formats and space division multiplexing to achieve substantial enhancement of the spectral efficiency \cite{Fontaine,Rademacher,Randel,PW}, while fiber remains a nonlinear medium leading to signal being strongly degraded by nonlinear impairments. In future systems, potentially using different types of fibers, nonlinear effects still will be one of the fundamental limiting factors. Therefore, compensation of the nonlinear signal distortions is a critical challenge for development of the next generation of communication systems operating at higher transmission rates. 

Up to date, non-linear impairment mitigation has been considered mostly for single mode fiber links with a number of techniques being proposed both in the optical and electronic domain \cite{OPC,regen,Radic1,Ip,DSP_Seb}. 
Most of the research in the field of electronic compensation has been focused on back-propagation algorithms that emulate optical transmission in the digital domain through an inverse fiber link (with reversed order segments and opposite sign parameters), realized either by means of a Split-Step Fourier method \cite{Ip,DSP_Seb} or Volterra Series Transfer Functions \cite{V1,V2,V3}. Despite various simplifications that have been proposed, both approaches are considered to be highly complex because they require multiple computational steps along the link and at least two samples per symbol. This has heavily discouraged any effort for future commercial deployment even for legacy SMF fiber systems. 

On the other hand, perturbation analysis of the Manakov equations has led to the development of efficient equalization methods that can mitigate accumulated intra-channel impairments in a single computational step and one sample-per-symbol \cite{Tao,Karlsson}. Central to this approach has been the identification of the perturbation coefficients that describe the interaction of each symbol with its preceding and succeeding symbols in the transmission channel\cite{Gao,Zhuge,Peng,quant}. For static connections and specific pulse shapes, such as Gaussian or sinc, the perturbation coefficients can be derived analytically and stored in a look-up table \cite{Tao,ECOC_Essiambre}. Since the total number of terms can be excessively high for dispersion un-managed links, where the channel memory is long, exploiting common symmetries and quantizing the coefficients are two of the techniques that have been recently proposed for complexity reduction. Furthermore, to achieve operation in reconfigurable network environments, an adaptive version of the method that uses training sequences and decision-directed least-mean squares algorithms has been introduced in \cite{McGill}. This enables to identify the perturbation coefficients before establishing any new connection and without prior knowledge of the corresponding transmission link parameters. The latter progress places the perturbation method in a broader context and signifies its practical importance in future optical networks.  

With this paper, we expand the application of perturbation-based nonlinear compensation in few mode fiber transmission systems by introducing a novel channel model, which captures nonlinear interplay between different modes in weak and strong coupling regime: Sparse Identification for Nonlinear Optical communication  systems: SINO method.  Contrary to the aforementioned approaches that deal only with the intra-channel nonlinearities, here, the proposed SINO method takes into account also the nonlinear interaction between the co-propagating modes by introducing additional inter-channel perturbation terms. The associated complexity scaling was addressed by adapting sparse identification method \cite{Kutz}, which makes use of the Lasso algorithm \cite{LASSO1,LASSO2}, thus, enabling computation of perturbation coefficients with inherent principal component analysis. The latter minimizes a mean squared error (MSE) estimation based on the training sequence and removing redundant predictors to improve model accuracy.
The method does not require knowledge of the transmission line and is applicable for multi-dimensional multichannel systems.

\section{Sparse identification  for nonlinear optical systems }
\subsection{Problem formulation}
\begin{figure*}
\begin{center}
\includegraphics[width=12cm]{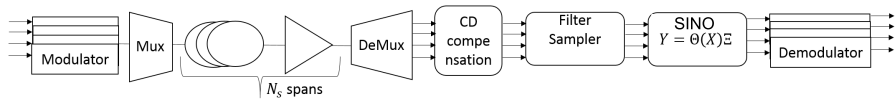}
\end{center}
\caption{  \label{Scheme} Scheme: SINO performs nonlinear MIMO equalization where the library is formed by nonlinear time-delayed function of the sampled symbols $\Theta(\mathbf{X})$ and the sparsity matrix $\Xi$, which contains information about nonlinear properties of the channel }
\end{figure*}

\begin{figure*}
\begin{center}
\includegraphics[width=12cm]{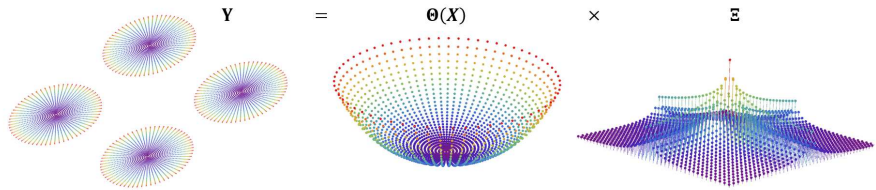}
\end{center}
\caption{  \label{F1} Schematic representation of the principle: the output signal $\mathbf{Y}$ is represented through nonlinear combination of input signal $\Theta(\mathbf{X})$ via sparse matrix $\Xi$}
\end{figure*}

\begin{figure*}
\begin{center}
\includegraphics[width=10cm]{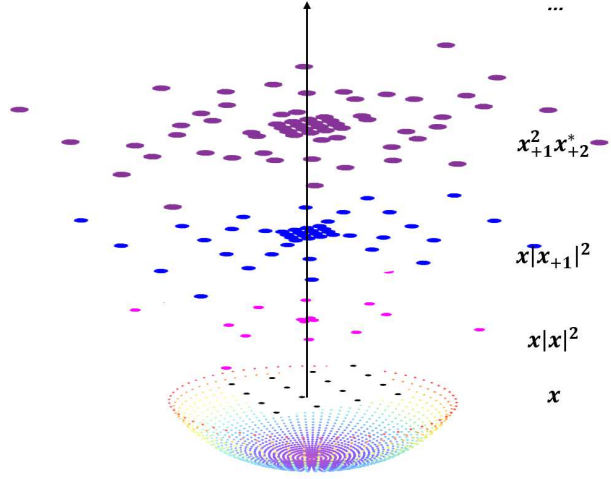}
\end{center}
\caption{  \label{F2} The library is constructed from the received symbols according to the channel model  Eq. \ref{MCM1_final}, i.e.  $\Theta(\mathbf{X})=[... x_k x_k|x_k|^2 ... x_{k+m}x_{k+n}x^*_{k+m+n}... ]$ (here 16 QAM is plotted for illustration).}
\end{figure*}

To demonstrate the application of the SINO method we consider the transmission of a $D$ spatial modes and two polarizations along a few mode fiber link of $N_s$ amplified spans, see Fig. \ref{Scheme}. At the receiver, a demultiplexer performs an initial separation of the spatial super-channel by projecting its tributaries onto a fixed mode basis. After coherent detection, matched filtering and down-sampling, the signals are fed into a linear MIMO equalizer, which compensates dispersion and remaining mode mixing effects. The transmission nonlinearities are treated separately by a non-linear MIMO processor (see Fig. \ref{F1}).  Unlike the linear equalizer, the processor creates a nonlinear matrix $\Theta(\mathbf{X})$ (Fig. \ref{F2}), which we call library, by forming mixing products of the input symbols. Depending on the channel memory the library will encompass different combinations - nonlinear mixing - of input symbols, thus, the library will have more elements that the original modulation format. See how library expands by different combinations of 16-QAM symbols in Fig. \ref{F2}: without memory in linear scenario the library has 16 symbols, taking into account nonlinearity (self-phase modulation) results in additional 16 symbols (formed from the original 16-QAM by a simple nonlinear transformation - $x_k|x_k|^2$, while adding memory expands the library further - for $M=1$ there are 48 new symbols formed as $x_k|x_{k+1}|^2$, further increase of memory expands the library even more.  Subsequently, it combines together the resulting terms through a coupling matrix $\Xi$, which contains information about the transmission line parameters and the pulse-shape of the signal. As symbols interact with different weights, the impact of some of these terms can be negligible, hence the matrix $\Xi$ is sparse and we can employ machine learning tools in identifying it. We may write: 

\begin{equation}\label{ML1}
\mathbf{Y}=\Theta (\mathbf{X})\Xi
\end{equation}
where $\mathbf{Y}$ is a vector of length $m$ representing the output signal. The nonlinear system of Eq. \ref{ML1} can be viewed as a representation of the output $\mathbf{Y}$ via a set of coordinates $\Xi$ on a complex functional space $\Theta(\mathbf{X})$. The sparsity of $\Xi$ allows us to simplify significantly the computational complexity of the nonlinear filtering process by removing redundant coefficients. As we will see below, this is an inherent characteristic of the lasso method, which has been adopted here for identifying the optimum $\Xi$. 


On the other hand, the exact form of the library matrix $\Theta(\mathbf{X})$ depends on the type of non-linearity the equalizer has to address. Given that in our case the fiber channel has memory we should expect each row of the matrix to involve not only elements, such as, $x(t_1)... x^2(t_1)...$, but also the non-instantaneous interactions between the symbols in different time slots, such us $|x(t_1)|^2x(t_2)$, therefore $\Theta(\mathbf{X})$ will be a dynamic matrix. Being able to identify its exact form is expected to improve significantly the accuracy and the convergence of the equalization process. This was achieved through a perturbation analysis on the Manakov equations that govern signal propagation along the FMF link and the derivation of a novel \emph{discrete-time multivariate channel model}, which is described in detail below. An important result of this derivation was the fact that we could analytically define the memory effects characterizing nonlinear interference within each channel and among the co-propagating modes and incorporate this dynamic behavior into $\Theta(\mathbf{X})$.

\subsection{SDM Channel Model}

For deriving the discrete-time channel model, and consequently the library matrix $\Theta (\mathbf{X})$, we assume that the propagation of an optical signal via an few mode optical fiber is governed by the Manakov equation for the $D$-modes \cite{SDM_eq,Mecozzi_SDM,Xiao}:
\begin{equation}\label{NLSE}
\frac{\partial U_p}{\partial z}=-\frac{\alpha}{2}U_p-\beta_1^p
\frac{\partial U_p}{\partial t}-i \frac{\beta_2^p}{2}
\frac{\partial^{2} U_p}{\partial^{2} t}+ i\kappa_{pp}\gamma_{pp}|U_p|^2 U_p +i\kappa_{pq}\sum^D_{q\neq p}\gamma_{pq}|U_q|^2U_p+
\eta_p(t,z),
\end{equation}
here deterministic distortions are described by  fiber losses $\alpha$, second-order dispersion $\beta_2$, and nonlinearity coefficient $\gamma$, whereas noise is zero-mean AWGN with variance $\langle \eta(z,t),\eta^*(z',t') \rangle=\frac{N_D}{L}\delta(z-z')\delta(t-t')$ where $N_D$ and $L$ are notations for noise spectral density and transmission length correspondingly. The formula captures both a) weak and b) strong coupling regimes, when a) $\kappa_{pp}=8/9, \kappa_{pq}=4/3$ and b) $\kappa_{pq}=\kappa_{pq}=8D/3/(2D+1)$. 

Given the expansion of the signal over pulses (i.e. the field component of polarization state $i$  $U_{pi}(t,0)=\sum_{k=-\infty}^{\infty}x_{kpi}f(t-kT))$,  after matched filter the continuous-time signal $U_p(t,L)=\{U_{p1},U_{p2}\}$
undergoes dispersion compensation and is sampled at $t=kT$: $Y_{kpi}(z)=P^{-1/2}\int dt \mathrm{D}[U_{pi}(t,z)]f(t-kT)$   results in a
discrete-time channel model:
\begin{equation}\label{NLSE}
Y_{kpi}'=\frac{\rho}{L}\eta_{kpi}+\varepsilon V[Y]_{kpi}, 
\end{equation}
with nonlinear potential
\[V[Y]_{kpi}=\sum_{j=1,2} \Psi_s(z)\sum_{m,n=-M}^{M}
(\varepsilon_{pp}Y_{k+n,pj}Y^*_{k+m+n,pj}\tilde{C}_{mn}^{pp}+\sum^D_{q\neq p}\varepsilon_{pq}Y_{k+n,qj}Y^*_{k+m+n,qj}\tilde{C}_{mn}^{pq})Y_{k+m,pi}\]
with AWGN noise term  having correlation  $\langle \eta_{kpi}(z),\eta_{k'p'i'}^*(z') \rangle=\delta(z-z')\delta_{kk'}\delta_{pp'}\delta_{ii'}$  and
where $\Psi_s(z)=e^{-\alpha mod(z,L_s)}$ is the signal power profile and $L_s$ is the span length. The coupling coefficients define the memory effects in the channel, it depends on pulseshape and system parameters:
\begin{equation}\label{Cmn}
C_{mn}^{pq}=i \int dz \Psi_s(z)  \int \int \int  d\omega  d\omega_1 d\omega_2 e^{i\omega_1(\beta_1^p-\beta_1^q) z+i(\omega_1^2+2\omega_1\omega)(\beta_2^p-\beta_2^q) z-i\omega_1\omega_2\beta_2^q z-i\omega_1mT-i\omega_2nT}\times
\end{equation}
\[ f^*(\omega)f(\omega_1+\omega)f(\omega_2+\omega)f^*(\omega_1+\omega_2+\omega) \]
Thus, we employ multiple scale analysis with small parameters: $ \varepsilon_{pp}=\kappa_{pp}\gamma_{pp} P z_d, \ \ \ \varepsilon_{pq}=\kappa_{pq}\gamma_{pq} P z_d, \ \ \  \rho=\sqrt{N_DB/P}$ with $B$ denoting signal bandwidth.
Assuming solution in the form: $Y=\sum_{n=0}^{\infty}\varepsilon^n Y_n$,
in the main order we have linear channel
with AWGN noise $\zeta_k$
having statistics $\langle \zeta_k,\zeta^*_m\rangle=\delta_{km}$:
\[
Y_{kpi}^{0}=X_{kpi}+\rho\zeta_{kpi}, \ \ \ \zeta_{kpi}=\int dz \eta_{kpi}(z)
\]
Then in the first order over nonlinearity  we have the nonlinear distortion:
\begin{equation}\label{MCM1_final}
Y_{kpi}^{1}=\sum_{m,n=-\infty}^\infty X_{k+n,pj}\Big(\varepsilon_{pp}C_{mn}^{pp}X_{k+m,pi}X_{k+n+m,pi}^* +\sum^D_{q\neq p}\varepsilon_{pq}C_{mn}^{pq}X_{k+m,qi}X_{k+n+m,qi}^* \Big)\end{equation}

Equation \ref{MCM1_final} reveals the memory property of the channel, which is that the interference between symbols decays exponentially with the symbol distance. Having exact knowledge of this property enables the development of a more efficient equalization process. 

The impact of simplifications on coupling matrix was studied in \cite{quant}, where $C_{mn}$ was approximated by a finite number of quantized levels and the degradation was estimated in terms of Q$^2$-factor.
 To calculate the latter one needs to calculate numerically a $(2M+1)^2$-number of 4-dimensional integrals (for a single channel case or $D(2M+1)^2$ for D-mode case). For a specific case of sinc-pulses single-channel WDM the problem has been reduced to 1-dimensional integrals \cite{ECOC_Essiambre} and has experimentally demonstrated the benefits of the model application for nonlinearity mitigation of intra-channel interference in multichannel-WDM.

In the next section we show how to obtain the matrix numerically using sparse identification and generalization of the model for higher nonlinearity.

\subsection{Stage1: Building library}
\begin{figure*}
\begin{center}
\includegraphics[width=12cm]{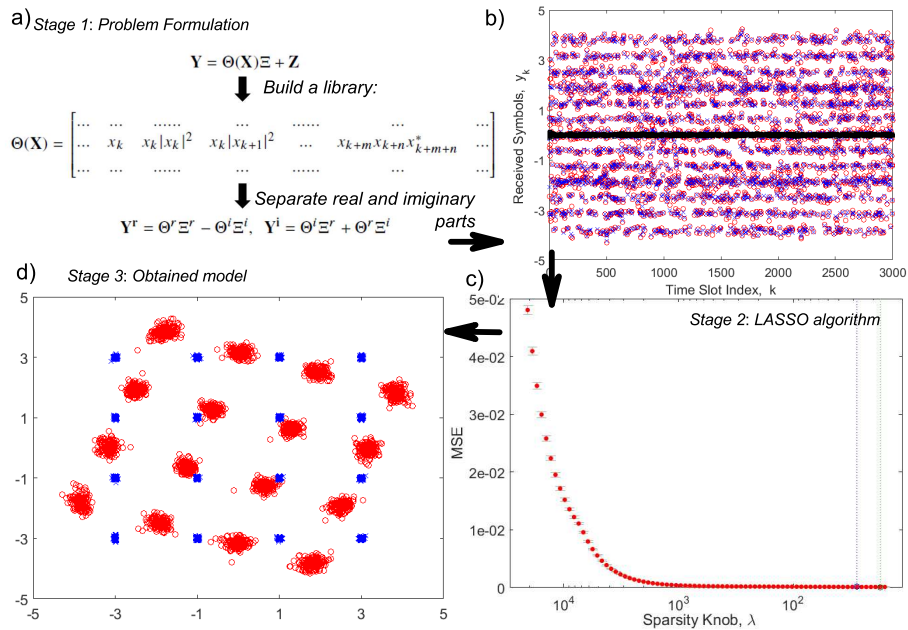}
\end{center}
\caption{  \label{Lasso} The figure shows the operating principle of our scheme. a) First we form the library matrix $\Theta(\mathbf{X})$. Then we apply sparse regression to identify the strength of interference between the components of $\Xi$. Here, the LASSO algorithm is applied which works  with real elements, therefore real and imaginary parts needs to be separated; b) the real part of the received symbol is plotted by red circles; c) LASSO identifies the matrix $\Xi$ by minimizing mean squared error as a function of optimization parameter $\lambda$; the resulted matrix $\Xi$ enables to establish a relation between output (red circles) and nonlinear combination of input symbols $\Theta$. Compare simulations (blue crosses) with the symbols obtained from the model (red circles), while difference between them is plotted in black squares; d) once the channel is fully identified it can be used for compensation of the non-linearity, compare distorted symbols in red and recovered in blue.}
\end{figure*}

As we know the type of nonlinearity that takes place we can construct the library (see Fig. \ref{Lasso}a) as follows
\begin{equation}\label{ML2}
\Theta(\mathbf{X}) = \begin{bmatrix} ... & ... & ... ... & ... & ...... & ... & ...          \\
       ... & x_k &x_k|x_k|^2 & x_k|x_{k+1}|^2 &... & x_{k+m}x_{k+n}x^*_{k+m+n}&...         \\[0.3em]
       ... & ... & ... ... & ... & ...... & ... & ... 
     \end{bmatrix}
\end{equation}

Also, as all vectors are complex we will separate real and imaginary parts (see Fig. \ref{Lasso}b):
\begin{equation}\label{ML2}
\mathbf{Y^r}=\Theta^r \Xi^r-\Theta^i \Xi^i, \ \ \mathbf{Y^i}=\Theta^i \Xi^r+\Theta^r \Xi^i
\end{equation}

Finally, we can apply this method directly to output signal $A(t)$, while it is very important to receive discrete-time representation by using appropriate filter, e.g. matched filter. The latter is particularly fit for identification of discrete-time model within information-theoretic treatment.

For real symmetric pulseshapes $f(\omega)=-f(\omega)=f^*(\omega)$ and $\beta_{1,2}^p=\beta^q_{1,2}$ we can reduce complexity by exploiting matrix symmetries: $C_{mn}=C_{nm}=C_{-m,-n}=-C^*_{-m\neq 0,n}$, which reflects in the appropriate reduced-dimensionality of the library $\Theta(\mathbf{X})$.

\subsection{Stage2: LASSO application}
As the problem is formulated, next we can apply any of the machine learning algorithms for a sparse solution of an overdetermined system. So that LASSO inherently performs principal component analysis, determining the minimum number of non-zero elements in $\Xi$, consequently, the minimum number of required operations. 
Once we have the function basis we can use various methods to calculate the sparse matrix $\Xi$.
In particular, here we applied least absolute shrinkage and selection operator (LASSO) \cite{LASSO1,LASSO2} using mean-squared error as a cost function.

As data and library are prepared we are looking to solve the problem to find a minimum:
\begin{equation}\label{Lasso1}
\min_{\beta_0,\beta} \Big( \frac{1}{2m}\sum_{i=1}^m(y_i-\beta_0-(\Theta^r \Xi^r-\Theta^i \Xi^i)_i^T\beta)^2+\lambda P_\alpha(\beta)\Big)
\end{equation}
where the penalty term:
\begin{equation}\label{Lasso2}
P_\alpha(\beta)=\frac{1-\alpha}{2}||\beta||^2_2+\alpha||\beta||_1
\end{equation}
In our simulations we applied the lasso fit with ten-fold cross validation (see Fig. \ref{Lasso}c).
The resulted matrix $\Xi$ comprises coefficients of the $C_{mn}$ matrix in Eq. \ref{Cmn} with simultaneous principle component analysis, which determines the minimum complexity operation. Once, the system is fully identified -the matrix $\Xi$ is determined, one can use it to compensate nonlinear impairments (see Fig. \ref{Lasso}d).

\section{Results and discussion}

The simulations parameters are summarized in Table 1. We investigated the transmission of a single wavelength spatial super-channel, along a 10-span FMF link of total length 10x100-km=1000-km. We considered $D$ spatial modes, each one of them including two polarization states, with $D$ taking values of 1, 3, and 6. On each of the 2D subchannels we launched a 4096-symbol, 16-QAM modulated stream of root-raised cosine pules (0.01 roll-off) at a symbol rate of 32GBaud and sampling rate of 16 samples-per-symbol (SpS). A FM-EDFA of 4.5 dB noise figure was considered after each span for compensating the propagation losses. Our focus was on investigating the equalization performance on the Kerr nonlinear effects only. Therefore, for the signal transmission we considered for every mode a Manakov-type of the propagation equation, see \ref{NLSE}, where the nonlinear effects are already averaged over all polarization states due to the randomly changing birefringence.  Also, for simplicity and without loss of generality, the modes belonged in the same group and they were strongly coupled, which implies a single nonlinear coefficient both for intra- and inter-modal effects (i.e. 1.4 1/W/km), and common group velocity and chromatic dispersion parameters. Finally, the performance was compared in terms of Q$^2$-factor which as calculated from the error vector magnitude (EVM), according to $Q^2 = 1/EVM^2$.

\begin{table*}[t]\label{Table}
\begin{center}
Table 1: System parameters
\\
\begin{tabular}{{ | c | c| r  |}}
    \hline
Bandwidth  & 32 GBaud \\
RRC Roll off & 0.01\\
  Distance & 10$\times$100 km    \\
  Noise Figure & 4.5 dB\\
  Attenuation coefficient & 0.2 dB/km
  \\
  CD & 17 ps/nm/km
  \\
  Nonlinearity & 1.4 1/W/km
  \\
\hline
\end{tabular}
\end{center}
\end{table*}

\begin{figure*}
\begin{center}
\includegraphics[width=12cm]{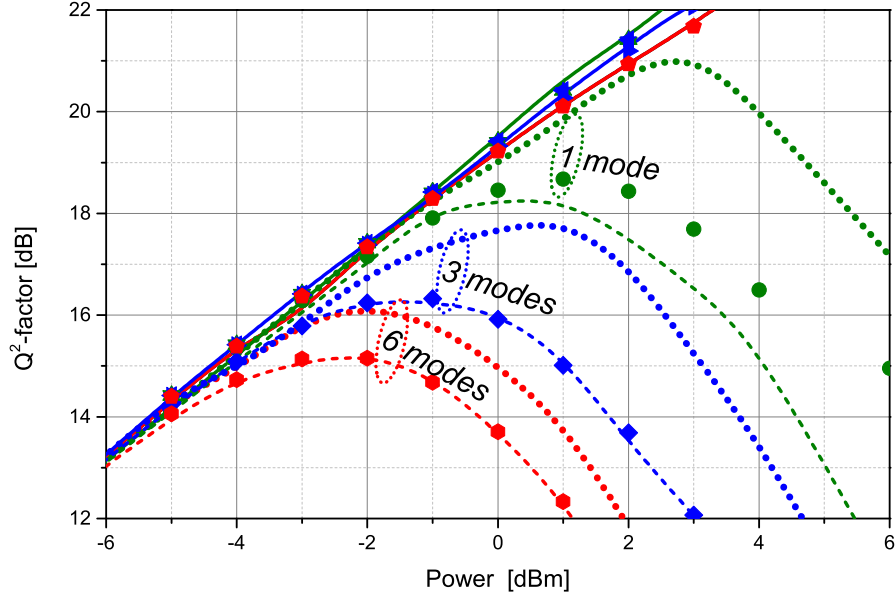}
\end{center}
\caption{  \label{Q} $Q^2$-factor vs. Launch power per spatial mode (i.e. average power of two pol. multiplexed sub-channels) using the proposed method (dotted) compared to ideal digital back propagation (solid) and linear compensation (dashed), for 1, 3 and 6 spatial modes (green, blue, red). Analytic expressions Eqs. \ref{SNR_1}-\ref{SNR_2} are plotted by filled symbols. }
\end{figure*}

\begin{figure*}
\begin{center}
\includegraphics[width=12cm]{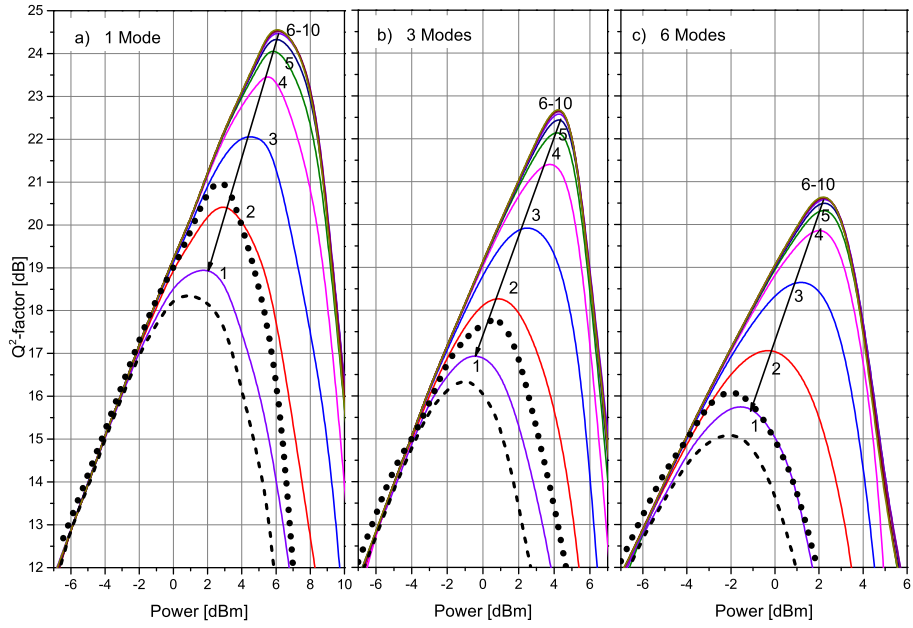}
\end{center}
\caption{  \label{DBP_steps} $Q^2$-factor vs. Launch power per spatial mode using the proposed method (dotted) compared to a symmetric digital back propagation algorithms with a different number of steps per span (1...10) and to linear compensation only (dashed) for 1, 3 and 6 spatial modes.   }
\end{figure*}
\begin{figure*}
\begin{center}
\includegraphics[width=12cm]{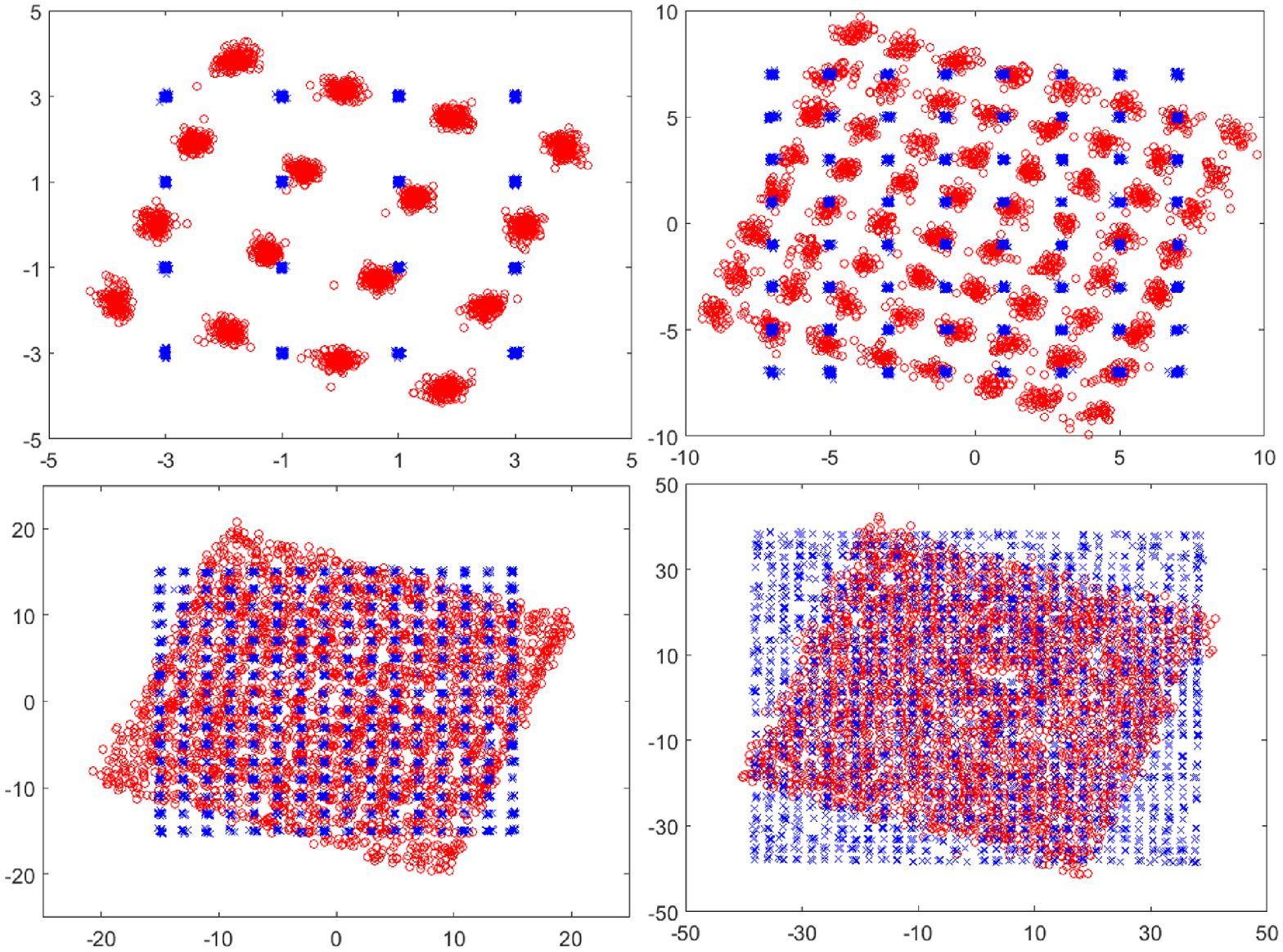}
\end{center}
\caption{\label{QAM} Uncompensated (red) and recovered (blue) constellations for 16-, 64-, 256-, and 1024-QAM after 1000km of single mode transmission with average launched power of $-3dBm$.}
\end{figure*}

In Fig. \ref{Q}, we compare the equalization performance of our proposed technique (dotted lines) with the case of equalizing linear impairments only (dashed lines), as well as, with the case of having ideal digital back propagation (DBP) (solid lines). 
Ideal DBP endures the full compensation of deterministic non-linearity and it is limited by nonlinear signal-noise interactions (here we ignored modal dispersion effects). Thus, the numerical estimations can be captured in a straightforward way by a simple analytic expression:
\begin{equation}\label{SNR_1}
SNR=\frac{S}{N+aS^3+bS^2N}, \ \ 
\end{equation}
the coefficient governing signal-signal interactions for ideal DBP scenario $a_{DBP}=0$, while for phase shift equalization
\begin{equation}\label{SNR_3}
a_{PSE}=2\sum_{q=1}^D \sum_{m,n\neq 0} \varepsilon_{pq}^2(2-\delta_{pq})|C_{m,n}^{pq}|^2
\end{equation}
while for the considered case $\beta_2^p=\beta_2^q, \beta_1^p=0, \gamma_{pq}=\gamma_{qq}$ the above expression can be easily simplified as
\[a_{PSE}=\frac{8}{27}(1+2(D-1))\frac{N_s\gamma^2L_{eff}\ln(\pi^2\beta_2L_{eff}B^2)}{\pi\beta_2B^2}, \]
(where $L_{eff}$ is an effective length). Since numerical estimations for the coupling matrix $\widetilde{C_{m,n}^{pq}}$ might deviate from the precise estimation, the coefficient transforms to 
\begin{equation}\label{SNR_4}
a_{SINO}=2\sum_{q=1}^D \sum_{m,n\neq 0} \varepsilon_{pq}^2(2-\delta_{pq})|C_{m,n}^{pq}-\widetilde{C_{m,n}^{pq}}|^2
\end{equation}
here $\delta_{pq}$ is the Kronecker symbol.
While the coefficient governing signal-noise interactions is given as
\begin{equation}\label{SNR_2}
b=\frac{4}{9}(1+2(D-1))\frac{N_s\gamma^2L_{eff}\ln(\pi^2\beta_2L_{eff}B^2)}{\pi\beta_2B^2}
\end{equation}

The aforementioned Eqs. \ref{SNR_1}-\ref{SNR_2} give a good estimate of the non-linearity impact for the different mode propagation scenarios (filled symbols in Fig. \ref{Q} by green, blue and red for 1, 3, 6- spatial modes - each including two polarizations states). For single mode propagation and at high launched powers, a deviation occurs suggesting the need of considering higher order nonlinear terms. 

The proposed nonlinear equalization scheme, results in an above 3 dB improvement for single mode transmission, which decreases to 2 and 1 dB for the 3 and 6 mode cases, respectively. The decrease in performance is attributed to the fact that we had considered only the 1st order terms in the nonlinear expansion of the perturbation model. Our proposed method can be straightforwardly applied to higher orders by augmenting the library matrix $\Theta$ with higher order term of $\mathbf{X}$, however, this is an issue to be addressed in a future work.

The performance of the proposed algorithm depends strongly on the calculation accuracy of the coupling matrix $\mathbf{C}$ through the evaluation of the sparse matrix $\Xi$. The calculations can be further improved by using more advanced methods for sparse matrix calculation than LASSO. Yet, in this simple configuration, which operated with a single sample per symbol and required just a single matrix multiplication, we were able to outperform conventional compensation techniques such as DBP. In Fig. \ref{DBP_steps} we have compared, in terms of received signal quality, our approach with the symmetric DBP algorithm which was solved for different number of steps per span and 2 samples per symbol. For single mode systems our method slightly outperforms the DBP of 2 steps per span. As the number of modes increases the performance difference is reversed in favor of DBP, so that for 6 modes our method compares to a DBP of single step per span. Even in that case, the complexity of the achieved equalization is significantly lower. 

Indeed, an important advantage of the proposed algorithm is that it ensures lowest complexity by evaluating the coupling matrix elements while removing the redundant terms. For example, in the simulated case the number of unique non-zero elements was 48 compared to the estimation of lowest complexity previous algorithm with adaptive filtering \cite{McGill} 96 components $(M-1)\log(M-1)+3M-1$ and non-optimized 324 $(2M+1)^2$ \cite{Tao}, where $M=\lfloor B^2 \beta_2 L/2\rfloor$ - estimation on channel memory ($\lfloor x \rfloor$ denotes a floor function of a variable $x$). Furthermore, the previous algorithms were developed only for compensation of inter-channel nonlinearities,  while in multichannel operation the complexity will increase proportionally to the number of channels. And the proposed algorithm with the inherent principal component analysis is crucial.

Finally, the SINO method is scalable to different modulation formats and signal powers as the matrix $\Xi$ remains unchanged, while power and modulation format influence are naturally incorporated in $\Theta$. Thus, once the matrix $\Xi$ has been identified during the establishment of a connection, and as long as the memory properties of the channel do not change, there is no need for retraining the algorithm and the same $\Xi$ can be used for any other modulation format  (see Fig. \ref{QAM}). This property makes the method extremely useful for flexible smart-grid network applications.

\section{Conclusions}
We have developed a low complexity machine learning based nonlinear impairment equalization scheme and demonstrated its successful performance in SDM transmission links achieving compensation of both inter- and intra- channel Kerr-based nonlinear effects. The method operates in one sample per symbol and in one computational step. It is adaptive, i.e. it does not require a knowledge of system parameters, and it is scalable to different power levels and modulation formats. Finally, although it has been developed for single wavelength spatial super-channels it can be straightforwardly expanded to multi-channel systems and to any other type of nonlinear impairment.
\\ 
This work has been supported by the EPSRC
project UNLOC EP/J017582/1. S. Sygletos acknowledges the support from the EU-FP7 INSPACE project under grant agreement N.619732. F. M. Ferreira is acknowledged for useful discussions 


\begin{thebibliography}{99}
\bibitem{DJR}  D. J. Richardson, "Filling the light pipe," Science \textbf{330}, 327-328 (2010).
\bibitem{CLEO} M. Sorokina, S. Sygletos, and S. K. Turitsyn, "Shannon capacity of nonlinear communication channels," in Conference on Lasers and Electro-Optics, OSA Technical Digest (2016) (Optical Society of America, 2016), paper SM3F.4.
\bibitem{Fontaine} 	N. K. Fontaine, R. Ryf, H. Chen, A. V. Benitez, B. Guan, R. Scott, B. Ercan, S. J. B. Yoo, L. E. Gruner-Nielsen, Y. Sun, R. Lingle, E. Antonio-Lopez, and R. Amezcua-Correa, "30$\times$30 MIMO transmission over 15 spatial modes," in Optical Fiber Communication Conference Post Deadline Papers, OSA Technical Digest (online) (Optical Society of America, 2015), paper Th5C.1.

\bibitem{Randel} 	S. Randel, R. Ryf, A. Gnauck, M. A. Mestre, C. Schmidt, R. Essiambre, P. Winzer, R. Delbue, P. Pupalaikis, A. Sureka, Y. Sun, X. Jiang, and R. Lingle, "Mode-multiplexed 6$\times$20-GBd QPSK transmission over 1200-km DGD-compensated few-mode fiber," in Optical Fiber Communication Conference, OSA Technical Digest (Optical Society of America, 2012), paper PDP5C.5.
\bibitem{Rademacher} 	G. Rademacher and K. Petermann, "Optimum capacity utilization in space-division multiplexed transmission systems with multimode fibers," in ECOC 2016, Th.2.P2.SC5.6.

\bibitem{PW}  P. J. Winzer,  "Scaling optical fiber networks: challenges and solutions," Optics and Photonics News, \textbf{26}, 28-35 (2015).
\bibitem{OPC} A. D. Ellis, M. E. McCarthy, M. A. Z. Al-Khateeb, and S. Sygletos, "Capacity limits of systems employing multiple optical phase conjugators," Opt. Express 23, 20381-20393 (2015).
\bibitem{regen} M. A. Sorokina and S. K. Turitsyn, "Regeneration limit of
classical Shannon capacity," Nat. Commun. 5, 3861 (2014).
\bibitem{Radic1}  E. Temprana, E. Myslivets, B.P.-P. Kuo, L. Liu, V. Ataie, N. Alic, S. Radic, "Overcoming Kerr-induced capacity limit in optical fiber transmission," Science  \textbf{348}, 1445-1448 (2015).
\bibitem{Ip} E. Ip, "Nonlinear compensation using backpropagation for polarization-multiplexed transmission," J. Lightwave Technol. 28(6), 939-951 (2010). 

\bibitem{DSP_Seb} S. J. Savory, "Digital coherent optical receivers: algorithms and subsystems," IEEE J. Sel. Top. Quantum Electron. 16(5), 1164-1179 (2010). 

 \bibitem{V1}  K. Peddanarappagari and M. Brandt-Pearce, "Volterra series transfer
function of single-mode fibers", J. Lightw. Technol, \textbf{15}(12), 2232-2241 (1997).
\bibitem{V2} M. Schetzen, "The Volterra and Wiener theories of nonlinear systems".
Malabar, Florida:Krieger, 2006.
\bibitem{V3}  
A. Amari, P. Ciblat, Y. Jaouen, "Fifth-order Volterra series based nonlinear equalizer for long-haul high data rate optical fiber communications," Asilomar Conference ACSSC (2014).

\bibitem{Tao} Z. Tao, L. Dou, W. Yan, L. Li, T. Hoshida, and J. C. Rasmussen, "Multiplier-free intrachannel nonlinearity compensating algorithm operating at symbol rate," J. Lightwave Technol. \textbf{29}(17), 2570-2576 (2011). 

\bibitem{Karlsson} P. Johannisson and M. Karlsson,  "Perturbation analysis of nonlinear
propagation in a strongly dispersive optical communication system," J.
Lightw. Technol.,  \textbf{31}(8),  1273-1282 (2013).

\bibitem{Gao} Y. Gao, J. C. Cartledge, A. S. Karar, S. S. Yam, M. O'Sullivan, C. Laperle, A. Borowiec, and K. Roberts, "Reducing the complexity of perturbation based nonlinearity pre-compensation using symmetric EDC and pulse shaping," Opt. Express 22(2), 1209-1219 (2014). 
\bibitem{Zhuge} Q. Zhuge, M. Reimer, A. Borowiec, M. O'Sullivan, and D. V. Plant, "Aggressive quantization on perturbation coefficients for nonlinear pre-distortion," in Optical Fiber Communications Conference and Exhibition (Optical Society of America, 2014), paper Th4D.7.
\bibitem{quant} Z. Li, W.-R. Peng, F. Zhu, and Y. Bai, "Optimum quantization of perturbation coefficients for
perturbative fiber nonlinearity mitigation," in \textit{Tech. Digest of European Conference on Optical Communication} paper We.1.3.4. (2014).
\bibitem{Peng} Z. Li, W. Peng, F. Zhu, and Y. Bai, "MMSE-based optimization of perturbation coefficients quantization for fiber nonlinearity mitigation," J. Lightwave Technol. 33(20), 4311-4317 (2015). 
\bibitem{ECOC_Essiambre}  A. Ghazisaeidi and R.-J.  Essiambre, "Calculation of coefficients of perturbative nonlinear pre-compensation for Nyquist pulses," in \textit{Tech. Digest of European Conference on Optical Communication paper} We.1.3.3. (2014).

\bibitem{McGill} M. Malekiha, I. Tselniker, and D. V. Plant, "Efficient nonlinear equalizer for intra-channel nonlinearity compensation for next generation agile and dynamically reconfigurable optical networks," Opt. Express 24, 4097-4108 (2016)











 
















\bibitem{Kutz} S. Brunton, J. Proctor and J. N. Kutz, Discovering governing equations from data by sparse identification of nonlinear dynamical systems, Proceedings of the National Academy of Sciences 113 (2016) 3932-3937.

\bibitem{LASSO1} R. Tibshirani, "Regression shrinkage and selection via the lasso," J. R. Stat. Soc. B \textbf{58}(1), 267-288 (1996).    
\bibitem{LASSO2} T.  Hastie, R.  Tibshirani, J. Friedman, "The elements of statistical learning," (Springer, New York), Vol. 2 (2009).


\bibitem{SDM_eq}	S. Mumtaz,  R.-J. Essiambre,  G. P. Agrawal, "Nonlinear propagation in multimode and multicore fibers: generalization of the Manakov equations," J. Lightw. Technol. 31(3), 398-406 (2013).

\bibitem{Mecozzi_SDM} A. Mecozzi, C. Antonelli, and M. Shtaif, "Nonlinear propagation in multi-mode fibers in the strong coupling regime," Opt. Express 20, 11673-11678 (2012).

\bibitem{Xiao} Y. Xiao, R. J. Essiambre, M. Desgroseilliers, A. M. Tulino, R. Ryf, S. Mumtaz, and G. P. Agrawal, "Theory of intermodal four-wave mixing with random linear mode coupling in few-mode fibers," Opt. Express 22, 32039-32059 (2014).



\end{thebibliography}
\end{document}